\documentclass[preprint,prd,aps,nofootinbib,showpacs]{revtex4}

\usepackage[dvips]{graphicx}
\usepackage{amsfonts, color, float, bbm, amsmath}
\usepackage{amssymb}

\def\beq {\begin{equation}}
\def\eeq {\end{equation}}
\def\bea {\begin{eqnarray}}
\def\eea {\end{eqnarray}}
\def\ni {\noindent}
\def\nn {\nonumber}

\def\lb {\left[ }
\def\rb {\right] }
\def\lc {\left\{ }
\def\rc {\right\} }

\def\rar {\rightarrow}

\def\cK {{\cal{K}}}
\def\cM {{\cal{M}}}

\def\cS {{\cal{S}}}

\def\d {\delta}

\def\g {\gamma}

\def\m {\mu}

\def\p {\pi}

\begin{document}

\title{Scalar resonances: scattering and production amplitudes}

\author{D. R. Boito}
\email[]{dboito@if.usp.br}
\author{M. R. Robilotta}
\email[]{robilotta@if.usp.br}

\affiliation{Instituto de F\'{\i}sica, Universidade de S\~{a}o Paulo,\\
C.P. 66318, 05315-970, S\~{a}o Paulo, SP, Brazil.}


\date{\today}

\begin{abstract}
Scattering and production amplitudes involving scalar resonances are
known, according to Watson's theorem, to share the same phase
$\delta(s)$.  We show that, at low energies, the production amplitude
is fully determined by the combination of $\delta(s)$ with another
phase $\omega(s)$, which describes intermediate two-meson propagation
and is theoretically unambiguous.  Our main result is a simple
and almost model independent expression, which generalizes the usual $K$-matrix unitarization procedure and is suited to be used in analyses
of production data involving scalar resonances.

\end{abstract}


\maketitle

In reactions such as $D^+ \rar \p^+ \p^- \p^+$ or $D^+ \rar \p^+ K^-
\p^+$, Dalitz plots of experimental data display a wealth of
resonating states (see for instance \cite{E791, Meadows}), indicating
that contributions from final state interactions (FSIs) to observed
quantities are important.  In order to disentangle basic interactions
from this kind of empirical information, a reliable theoretical description of the
final three-meson system is required.  As the full treatment of
this problem is rather involved,  one usually resorts to an
approximation, which consists in assuming that one of the final mesons
acts as a spectator. The limitations of this approximation are not
well established and the issue is being debated \cite{Meadows,
Caprini}.  In this {\it quasi two-body} approach, theoretical work
becomes much simpler, since one deals with two-body FSIs directly
related with elastic scattering.  This idea underlies Watson's theorem
\cite{Watson}, which was produced more than fifty years ago and states
that the same phase $\d(s)$ occurs in both scattering and production
amplitudes.  Important as it is, the theorem does not determine how
scattering information is to be used in the interpretation of
production data and one finds a multitude of mutually contradictory
prescriptions in the literature.  In particular, there are
parametrized production expressions based on either $(\sin \d)$
\cite{BediagaMiranda, Babar} or $(\cos \d)$ \cite{Achasov, Pennington,
CLEO, Furman}. In other cases, expressions used in data analyses
bear no connection with the scattering phase. For instance, the
pioneering study of the $\sigma(500)$ scalar resonance in the decay
$D^+\to \pi^+ \pi^-\pi^+$ \cite{E791} was based on a trial amplitude
of the form
\beq
\mathcal{A} = a_{nr} \mathcal{A}_{nr} + \sum_n a_n \mathcal{A}_n^{res},
\label{iso}
\eeq
where $nr$ stands for a non-resonant background, the
$\mathcal{A}_n^{res}$ represent the contributions of various
resonances and the complex weights $a_i$ are determined from fitting
to the data. The main ingredient of the $\mathcal{A}_n^{res}$
amplitudes is  a relativistic Breit-Wigner function, aimed at
describing resonance propagation and decay. This  expression makes
no use of the $\pi\pi$ scattering amplitude. 
An obvious problem with this loose adoption of trial
functions to interpret empirical data is that
 information about the position of the
resonance pole in the complex energy plane becomes contaminated by
unreliable assumptions.

We tackle the scattering-production problem in the case of a scalar
resonance.  In order
to simplify the notation, we refrain from writing angular momentum or
isospin labels explicitly in amplitudes and phases.  We assume that
the scalar resonance $R$ is coupled to a pair of pseudoscalar mesons
$P_a P_b$ which, in practice, represents systems such as $\p \p$, $\p
K$, $K\bar K$, in the presence of $\sigma(500)$, $\kappa(800)$ and
$f_0(980)$ resonances.  The meson masses are $\m_a$ and $\m_b$. As we
are mostly interested in relatively low-energy processes involving pions and
kaons, in this work
we follow both the conceptual and mathematical descriptions of resonances
proposed in Ref. \cite{Pich}.

The amplitude $T$ describing the elastic process 
$P_a P_b \rar P_a P_b$ must respect unitarity, and below the first inelastic threshold, it can
always be written as  \footnote{The amplitude $T$ is relativistic and we employ the conventions of Refs. \cite{Achasov} and \cite{Robilotta}.}
\beq
T(s) = \cS \, \frac{16 \p}{\rho(s)} \,  \sin \delta(s) \, e^{i \delta(s)} ,
\label{e.1}
\eeq

\ni
where $\cS$ is a statistical factor such that 
$P_a = P_b \rar \cS=2$, $P_a \neq P_b \rar \cS=1$, and 
$\rho(s) = \sqrt{[1-(\m_a + \m_b)^2/s][1-(\m_a - \m_b)^2/s]}$.
Unitarity is implemented automatically in this parametrization, 
by means of the real function $\delta(s)$, which
encompasses the full dynamical content of the interaction.

\begin{figure}[!ht]
\begin{center}
\includegraphics[width=0.9\columnwidth,angle=0]{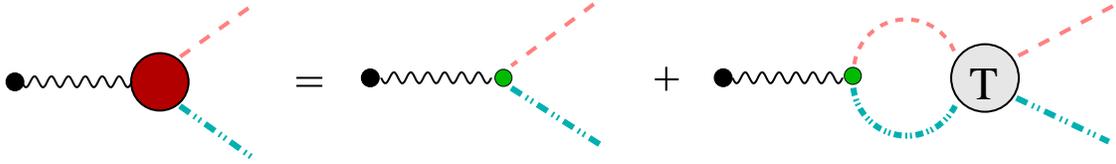}
\caption{{ Resonance (wavy line) propagation and decay into mesons $P_a$ (dashed line) and $P_b$ (dot-dashed line). $T$ stands for the the unitary scattering  amplitude, Eq. (\ref{e.1}). } } 
\label{fig:SDecay}
\end{center}      
\end{figure}

In decays of heavy mesons, such as $D$ or $B$, a part of the 
width may be due to the direct production of a resonance  at the weak vertex.
When this happens, FSIs become important and the interpolation between the  decay vertex and the observed $|{P_aP_b}\rangle$ state is described by the 
subset of diagrams shown in Fig. \ref{fig:SDecay}, involving both the {\em bare}
resonance propagator and the {\em unitarized} elastic scattering amplitude \cite{Achasov,CLEO}. This subset of Feynman diagrams is represented by the function
$\Pi(s)$ and, for the sake of conciseness, referred to as  {\it production subamplitude}.

At low energies, $\Pi(s)$ is designed to replace the Breit-Wigner
function associated with the scalar resonance in Eq. (\ref{iso}) and
has the advantage of exhibiting a clear relation to the scattering
amplitude.  We demonstrate, in the sequence, that it can be expressed
in terms of the elastic phase $\delta(s)$ as
\beq
\mathrm{\bf{form \;\; 1:}}\;\;\; \Pi(s)= g\;\frac{\cos \delta(s)}{m_R^2 - s} \,
\lb 1 + \frac{\tan \delta(s)}{\tan \omega(s)} \rb \, e^{i \delta(s)},
\label{e.2}
\eeq

\ni 
where $g$ is the resonance-mesons coupling constant, $m_R$ is the 
nominal resonance mass, such that $\delta(s=m_R^2)=\pi/2$,  
and $\omega(s)$ is a meson loop phase given by
\beq
\tan \omega(s) =  \frac{\pi \rho(s)}{\Re [L(s) - L(m_R^2)   ]},
\label{e.3}
\eeq

\ni where,

\beq
\Re L(s) = \rho(s) \log \lb \frac{\sqrt{s - (\mu_a -\mu_b)^2} - \sqrt{s - (\mu_a +\mu_b)^2 }  }{\sqrt{s - (\mu_a -\mu_b)^2} + \sqrt{s - (\mu_a +\mu_b)^2 }}\rb - \frac{\mu_b^2 -\mu_a^2}{s}\log\left( \frac{\mu_b}{\mu_a}\right).
\eeq

This is our main result.
Simple manipulations allow it to be recast in two alternative forms,
namely
\bea
\mathrm{\bf{form \;\; 2:}}
&& \Pi(s)= g\;\frac{\cos \delta(s)}{\cM^2(s) - s} \, e^{i \delta(s)},
\label{form1}\\[3mm]
\mathrm{\bf{form \;\; 3:}}
&& \Pi(s)= g\;\frac{\sin \delta(s)}{m_R\, \Gamma(s)} \, e^{i \delta(s)},
\label{form2}
\eea

\ni 
where $\cM(s)$ is a running mass and $\Gamma(s)$ is a running width.
It is important to note that these two functions are not independent,
since both of them are unambiguously determined by the phase $\omega(s)$.

\begin{figure*}[!ht]
\begin{center}
\includegraphics[width=1.0\columnwidth,angle=0]{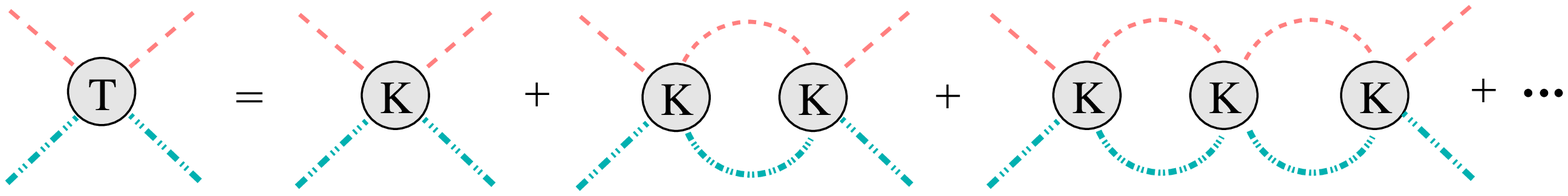}
\caption{{Geometrical series for the elastic unitary $P_aP_b$ scattering $T$-matrix. $K$ stands for the kernel, Eq. (\ref{2.6}).  } } 
\label{fig:TUnitaria}
\end{center}      
\end{figure*}

We now demonstrate the results given by
Eqs. (\ref{e.2})-(\ref{form2}).  In the spirit of the Bethe-Salpeter
equation, the dynamical content of a {\it unitarized} elastic
scattering amplitude can be realized as a series involving kernels
and propagators.  This idea is represented in
Fig. \ref{fig:TUnitaria}, where the iteration of the kernel $\cK$ by
means of a two-body mesonic Green's function yields the full amplitude $T$.

In the case of low-energy meson-meson scattering, the problem 
is much simplified by the fact that the two-body propagator  
can be described by a renormalized\footnote{The mesonic loop integral contains
an infinite contribution, which is eliminated by means of a couterterm, chosen
so that one has $\delta=\pi/2$ at a point $s=s_0$. As $s_0$ can be determined
by measurements of the phase shift, for convenience, we write $s_0\equiv m_R^2$. Further details can be found in Ref. \cite{Robilotta}, especially its appendix B.} analytic loop function, which we denote by 
$\bar \Omega(s)$.
In order to construct it, we note that the finite part of the 
two-meson propagator is given by a function $L(s)$ such
that 
\begin{align}
\bullet &\,\, \mbox{for}\,\, 0<s<(\mu_b - \m_a)^2 \nn \\
&L(s) = \rho(s) \log \lb \frac{\sqrt{(\mu_a +\mu_b)^2-s} + \sqrt{(\mu_a -\mu_b)^2 -s}  }{\sqrt{(\mu_a +\mu_b)^2-s} - \sqrt{(\mu_a -\mu_b)^2 -s }}\rb  -\frac{(\mu_b^2 -\mu_a^2)}{s}\log\left( \frac{\mu_b}{\mu_a}\right) , \\
 \bullet &\,\, \mbox{for}\,\,(\mu_b - \m_a)^2< s< (\m_b+\m_a)^2 \nn \\
 &L(s) = \frac{\sqrt{-\lambda(s)}}{s}
 \left\{ \tan^{-1}\left[  \frac{\m_a^2 + \m_b^2 -s}
 {\sqrt{-\lambda(s)}} \right] 
 -\frac{\pi}{2}  \right\} 
   - \frac{(\m_b^2- \m_a^2)}{s}\log \left(\frac{\m_b}{\m_a} \right),\\
\bullet &\,\, \mbox{for}\,\, s>(\mu_b + \m_a)^2 \nn \\
&L(s) = \rho(s) \log \lb \frac{\sqrt{s - (\mu_a -\mu_b)^2} - \sqrt{s - (\mu_a +\mu_b)^2 }  }{\sqrt{s - (\mu_a -\mu_b)^2} + \sqrt{s - (\mu_a +\mu_b)^2 }}\rb  - \frac{(\mu_b^2 -\mu_a^2)}{s}\log\left( \frac{\mu_b}{\mu_a}\right) + i \pi \rho(s),
\end{align}
\ni where  $\lambda(s) = (s - \mu_a^2 -\mu_b^2)^2 - 4\mu_a^2\mu_b^2$.
The expression for $L(s)$ is simplified if $\mu_a=\mu_b$ and can
be found  in \cite{Robilotta}.
Above threshold, the function $\bar \Omega(s)$ is written as
\bea
\bar \Omega(s) &=& -\frac{1}{\cS \, 16\p^2} 
\lc \Re \lb L(s) - L(m_R^2) \rb
+ i\, \Im  L(s)  \rc
\nn\\
&\equiv& \bar R(s) + iI(s)\label{omega}
\eea
\ni
and, by construction, $\bar R(m_R^2)=0$.
The imaginary component is particularly simple and reads
$I(s)=-\rho(s)/(\cS\,16\p)$.
The phase $\omega(s)$ entering Eq. (\ref{e.3}) is defined as 
\beq
\tan \omega(s) \equiv I(s)/\bar R(s) \;.
\label{eo}
\eeq

The representation of $T$ given in Fig.2
corresponds to a geometrical series\footnote{The fact that this series
can be summed was shown by Oller \& Oset in Ref. \cite{Oller}. It is
a particular property of the interaction kernel involving pions and kaons.} and yields
\beq
T = \cK + \cK (-\bar \Omega)  \cK 
+ \cK  (-\bar \Omega)  \cK  (-\bar \Omega)\cK + \cdots
= \frac{\cK}{1 + \bar \Omega \cK} .
\label{2.5}
\eeq

In a theory containing a resonance as an explicit degree of freedom,
it is convenient to factorize its $s$-channel  denominator 
in the kernel $\cK$ and we write 
\beq
\cK \equiv \frac{\g^2(s)}{m_R^2 -s}\;,
\label{2.6}
\eeq
\ni
where $\g(s)$ is a function satisfying the condition 
$\g(m_R^2)\neq 0$.
It is important to stress that Eq. (\ref{2.6}) corresponds just to a definition, which is completely
independent of dynamical assumptions.
The specific form of $\g(s)$ for $\p\p$ scattering in the framework of a $SU(2)$ implementation of chiral symmetry was discussed in Ref. \cite{Robilotta}, but it may be ignored in
the present work. 
Using Eq.(\ref{2.6}) into Eq.(\ref{2.5}), one finds
\beq
T= \cS \,\frac{16\p}{\rho(s)}\; 
\frac{m_R\,\Gamma(s)}{\cM^2(s) - s -i\,m_R\, \Gamma(s)} \;,
\label{2.7}
\eeq 

\ni
where we have defined a running mass and a running width by
\bea
&& \cM^2(s) \equiv m_R^2 + \g^2(s) \,\bar R(s) \;,
\label{2.8}\\[2mm]
&& m_R\,\Gamma(s) \equiv \frac{\g^2(s)\,\rho(s)}{\cS \, 16 \p} 
\label{2.9}
\eea
\ni and, by  construction, $\cM^2(m_R^2)=m_R^2$. 
Comparing this result with Eq.(\ref{e.1}), one has
\beq
\tan \delta(s)= \frac{m_R \,\Gamma(s)}{\cM^2(s) -s} \;.
\label{2.10}
\eeq

On the other hand, the elimination of $\g^2(s)$ from Eqs. (\ref{2.8})
and (\ref{2.9}) makes the constraint between the running 
functions evident and allows one to write
\beq
\frac{1}{\cM^2(s) -s} = \frac{1}{m_R^2 -s}\left( 1 + \frac{\tan \delta(s)}{\tan \omega(s)} \right),
\label{eq:tantan}
\eeq

\ni
after using Eq.(\ref{eo}). 
This result plays an important role in our derivation, because it
indicates that the running mass is completely determined just by the 
information contained in the empirical function $\delta(s)$ and the 
theoretically reliable phase $\omega(s)$.


The production subamplitude $\Pi(s)$  can be evaluated directly from Fig. \ref{fig:SDecay}
and reads
\beq
\Pi(s) = \frac{g}{m_R^2-s} 
\lb 1 + (-\bar\Omega) \frac{\cK}{1 + \bar\Omega\cK} \rb
= \frac{g}{m_R^2-s} 
\lb \frac{1}{1 + \bar\Omega\cK} \rb .
\label{2.11}
\eeq
\ni
As stressed by  Pennington \cite{Pennington}, this shows
that production and scattering  share a ``universal
denominator'', namely $(1 + \bar\Omega\, \cK)$, which imposes the Watson
phase to the production subamplitude and transmits the physical poles
from one amplitude to the other.  Using Eqs.
(\ref{2.5}) and (\ref{2.6}), one finds Eq. (\ref{form2})
\beq
\Pi(s) = \frac{  g  }{\g^2(s)} \lb \frac{\cK}{1 + \bar \Omega\,\cK} \rb
= \frac{g\, T}{\g^2(s)}  
= \frac{g}{m_R\Gamma(s)} \sin \delta(s) e^{i \delta(s)},
\label{2.12}
\eeq

\ni which is form 3. Using Eq. (\ref{2.10}) it can
be transformed into form 2, Eq. (\ref{form1}). Finally, Eq.
(\ref{eq:tantan}) yields form 1, Eq ({\ref{e.2}}). This  concludes our
demonstration.

Another interesting relationship that can be derived from Eqs. (\ref{e.2}) and (\ref{2.12}) is 
\beq
\cK = \mathcal{S}\frac{ 16\pi}{\rho(s)}\frac{\tan \delta}{ \left(1+ \tan\delta/\tan\omega \right)    },
\label{Komega}
\eeq
showing that the kernel is determined just by $\delta$ and $\omega$.

Results (\ref{e.2})-(\ref{form2}) extend previous ones available in
the literature. Our unitarization, based on the $\bar \Omega$ two-body
Green's function, generalises the usual $K$-matrix approach
\cite{Achasov,Black}. In the latter, the particles propagating inside
the loop are taken to be on-shell, which amounts to ignoring the real
part of $\bar \Omega(s)$: making $\bar R\to 0$ in Eq. (\ref{omega}),
one has $\cot \omega \to 0$ and $\cM^2\to m_R^2$.  The
deviation of form 1 from the $K$-matrix result is quantified by the
factor $(1 + \tan\delta/\tan\omega)$.


\begin{figure}[!ht]
\begin{center}
\includegraphics[width=0.7\columnwidth,angle=0]{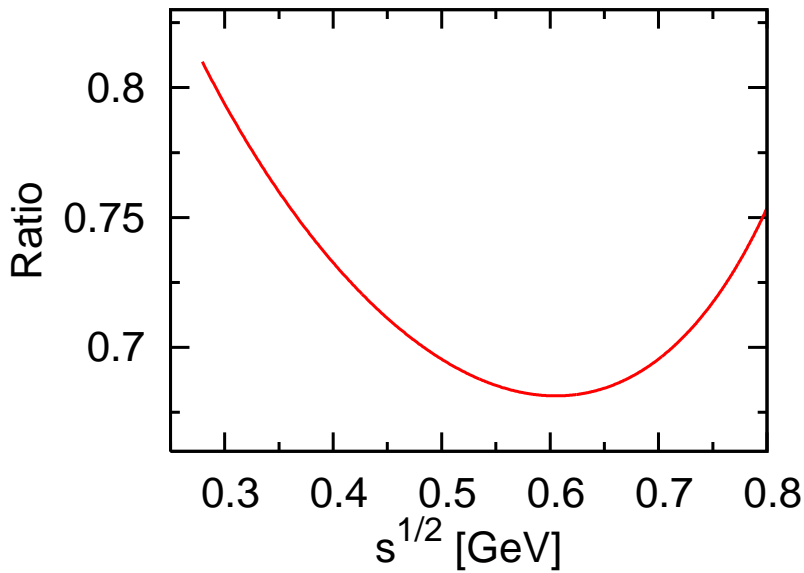}
\caption{{Ratio  of the result from the $K$-matrix approximation to our Eq. (\ref{e.2}), given by $(1+\tan\delta_0^0/\tan\omega)^{-1}$, for the $\pi\pi$ system. } } 
\label{fig:NewTerm}
\end{center}      
\end{figure}

\begin{figure}[!ht]
\begin{center}
\includegraphics[width=0.7\columnwidth,angle=0]{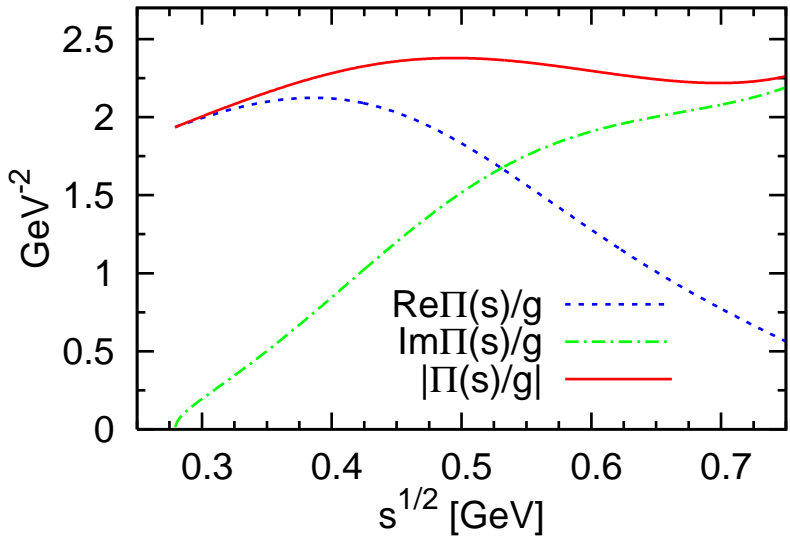}
\caption{{Result for $\Pi(s)/g$ obtained by means of Eq. (\ref{e.2}) using $\delta_0^0(s)$ from \cite{CGL}. } } 
\label{fig:PiCGL}
\end{center}      
\end{figure}

In order to produce a feeling for the numerical importance of this
discrepancy, we concentrate on the $\p \p$ system and adopt the
parametrization for the scalar-isoscalar phase shift $\delta_0^0(s)$
quoted by Colangelo, Gasser and Leutwyler (CGL) \cite{CGL}. We compute
$\Pi(s)/g$ using Eq. (\ref{e.2}) and identifying $m_R$ with their
$\sqrt{s_0}=0.846$ GeV, which corresponds to
$\delta_0^0(s_0)=\pi/2$. We show in Fig. \ref{fig:NewTerm} the ratio of the result obtained in a $K$-matrix approach to ours and note that the former is typically $20-30\%$ smaller in the range
$2\mu_{\pi} <\sqrt{s}<0.8$ GeV.  In Fig. \ref{fig:PiCGL} we show results for the
real and imaginary parts of $\Pi(s)/g$, as well as its modulus. The
profile of $|\Pi(s)/g|$ has a broad peak-like structure close to the
$\sigma(500)$ pole position\footnote{The result for the $\sigma$ pole
position of \cite{CGL} is not far from the latest and more precise
determination \cite{Leutwyler} $\sqrt{s_\sigma} = 441^{+16}_{-8} - i
272^{+9}_{-12.5}$ MeV.} quoted in \cite{CGL} namely, $\sqrt{s_\sigma}\approx
0.475 - i 0.290$ GeV.  This agrees  with both empirical peaks in
production experiments \cite{E791,BES,CLEO} and theoretical works dealing with
pionic FSIs based on Chiral Perturbation Theory
\cite{MeissnerOller, MeissnerGardner}.

In summary, we have presented a prescription that allows the
production subamplitude of a low-energy scalar resonance to be
determined from the elastic scattering phase, supplemented by another
one, describing a two-meson Green's function. This prescription has a
number of nice features, namely:


1. It has very little model dependence. As far as dynamics is
concerned, the only assumption made is that the kernel entering the
Bethe-Salpeter structure shown in Fig. \ref{fig:TUnitaria} can be considered as
point-like at low-energies. This assumption, which is standard in the
study of mesonic systems by means of effective theories, allows one to
use the rather well known bubble-like two-meson propagator. 

2. It contains the $K$-matrix approach as a particular case. In
that approach, corrections from the meson loops to the resonance mass
are neglecteded. However, as shown in Fig. \ref{fig:NewTerm} for the $\pi\pi$
system, the loss of accuracy induced by this simplification is of the
order of 30\% in the region of physical interest.

3. It gives rise to a bump which is also visible in experiments for the $\pi\pi$ case.

4. It indicates, by means of alternative forms 2 and 3, that
expressions involving both ($\sin \delta$) and ($\cos \delta$) are
acceptable, {\it provided the coherent running mass and width are used.} In
case other forms for these functions are adopted, consistency is lost
and results may go astray.

5. It shows that precise measurements of  both scattering and
production amplitudes can produce direct information about the
meson-meson interaction kernel. The effects of the unitarization
procedure, encoded in the $\bar \Omega$ function, can be separated
from the low-energy dynamics of the scattering process, which is
represented by the function $\gamma^2(s)$ in
Eq. ({\ref{2.6}}).  Using Eq. (\ref{Komega}) one learns that  this function can be determined from $\delta$ and $\omega$. This is important
because $\gamma^2(s)$ shapes the denominator of Eq. (\ref{2.7}) and plays
a crucial role in the extraction of resonance parameters by means of
pole extrapolation. Last, but not least, Eq. (\ref{Komega}) yields directly $\cK$ in terms of
$\delta$ and $\omega$. 

6. It complements Watson's theorem, by showing that scattering and
production amplitudes share not only a phase, but also neat
information about the meson-meson interaction kernel.

7. It can be tested experimentally by simply replacing a Breit-Wigner
function with $\Pi(s)$ in Eq. (\ref{iso}).

A more detailed  analysis of production data involving $\pi\pi$ and $\pi K$ resonances using  the formalism developed here is in progress.

\begin{acknowledgments}
We thank Dr. Ign\'acio Bediaga for discussions and information about empirical data. DRB thanks the
hospitality of LPNHE - Laboratoire de Physique Nucl\'eaire et de
Hautes Energies, at Universit\'e P. \& M. Curie - Paris VI,  where part of this work
was performed,  and especially Drs. Beno\^it Loiseau and Bruno El-Bennich. This work is supported by FAPESP (04/11154-0)
(Brazilian Agency).
\end{acknowledgments}


\end{document}